\documentclass[a4paper]{jpconf}
\usepackage{graphicx}
\begin{document}
\title{Testing models of extragalactic $\gamma$-ray propagation \\ using observations of extreme blazars \\ in $GeV$ and $TeV$ energy ranges}

\author{T A Dzhatdoev$^{1,*}$, E V Khalikov$^{1,**}$, A P Kircheva$^{2,1}$ \\ and A A Lyukshin$^{2}$}

\address{$^1$ Federal State Budget Educational Institution of Higher Education M.V. Lomonosov Moscow State University, Skobeltsyn Institute of Nuclear Physics (SINP MSU), 1(2), Leninskie gory, GSP-1, Moscow 119991, Russian Federation}
\address{$^2$ Federal State Budget Educational Institution of Higher Education M.V. Lomonosov Moscow State University, Department of Physics, 1(2), Leninskie gory, GSP-1, Moscow 119991, Russian Federation}
\ead{$^{*}$timur1606@gmail.com}
\ead{$^{**}$nanti93@mail.ru}

\begin{abstract}
We briefly review contemporary extragalactic $\gamma$-ray propagation models. It is shown that the Extragalactic Magnetic Field (EGMF) strength and structure are poorly known. Strict lower limits on the EGMF strength in voids are of order $10^{-17}-10^{-20}$ $G$, thus allowing a substantial contribution of a secondary component generated by electromagnetic cascades to the observable spectrum. We show that this ``electromagnetic cascade model'' is supported by data from imaging Cherenkov telescopes and the Fermi LAT detector.
\end{abstract}

\section{Introduction}

Observations made with contemporary imaging Cherenkov telescope arrays may be used to study fundamental processes in which very high energy (VHE, $E>$100 $GeV$) $\gamma$-rays are involved. Indeed, the column density of low-energy photons of the extragalactic background light (EBL) and the cosmic microwave background (CMB) over cosmological distances (redshift $z>$0.1) may appear so high that the probability of the $\gamma\gamma\rightarrow e^{+}e^{-}$ process becomes considerable \cite{nik62}--\cite{gou67}. Notwithstanding the large period of time that passed since \cite{nik62}--\cite{gou67} were published, the vast majority of works on extragalactic $\gamma$-ray propagation adhered to the same approach which accounts for the absorption of primary photons as the only process that changes the shape of the primary (intrinsic) source spectrum (the corresponding astrophysical model is hereafter referred to as the ``absorption-only model''). The only qualitative improvement of the state-of-art absorption-only model since 1962 is the inclusion of adiabatic losses.

In the present paper we consider other models of extragalactic gamma-ray propagation that go beyond the absorption-only model. Section 2 contains some contemporary constraints on the strength and structure of the Extragalactic Magnetic Field (EGMF) in voids of the Large Scale Structure. Section 3 briefly reviews the main astrophysical processes that may induce new effects in the spectral, angular and temporal distributions of observable $\gamma$-rays. Observational results in the $GeV$ energy region discussed in these sections were obtained with the Fermi LAT instrument \cite{atw09}. In sections 4--5 we present the  electromagnetic and hadronic cascade models that may better explain current data.

\section{Constraints on the Extragalactic Magnetic Field strength}

EGMF affects the motion of electrons and positrons (hereafter simply ``electrons'') produced by primary $\gamma$-rays on EBL and CMB. Therefore, secondary (cascade) $\gamma$-rays get deflected and delayed; thus, the observable $\gamma$-ray spectrum is dependent on the EGMF properties.

\begin{figure}[h]
\includegraphics[width=17pc]{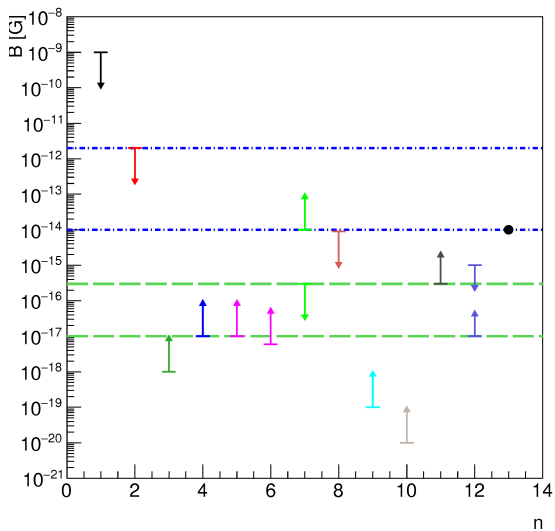}\hspace{2pc}%
\begin{minipage}[b]{19pc}\caption{\label{fig1} Constraints on the EGMF strength in voids.
$n$ denotes the number of corresponding work.
$n$=1 --- \cite{bla99}; $n$=2 --- \cite{dol05}; $n$=3 --- \cite{der11}; $n$=4 --- \cite{tay11}; $n$=5 --- \cite{vov12}, the case of the EBL model \cite{fra08}; $n$=6 --- \cite{vov12}, the case of the EBL lower limit from direct source counts; $n$=7 --- \cite{abr14} $n$=8 --- \cite{pro12} $n$=9 --- \cite{fin15} $n$=10 --- \cite{tak12} $n$=11 --- \cite{ner10} $n$=12 --- \cite{che15} $n$=13 --- \cite{tas14}. Blue lines show the range of EGMF strength values for which absorption-only model is accepatble, olive lines contain the acceptable region for EM cascade models.}
\end{minipage}
\end{figure}

EGMF strength and structure are poorly known. Still, some constraints on the strength of EGMF exist; most of them are heavily model-dependent. These constraints are shown in figure \ref{fig1}; this figure is an update of figure 4 published in \cite{dzh15}. Upper bounds on the EGMF strength $B<10^{-9}$ $G$ were put in \cite{bla99}. Other and more restrictive upper bounds  ($\approx2 \cdot 10^{-12}$ $G$), obtained from cosmological simulations, were introduced in \cite{dol05}. Several authors \cite{der11}--\cite{fin15} found that in case of zero EGMF the expected intensity in the $GeV$ region of some blazar spectra would exceed the observed one due to the contribution of the cascade component, and put more or less robust constraints on the $B$ value: $B>3 \cdot 10^{-16}$ $G$ in \cite{ner10}, $B > 10^{-18} - 10^{-17}$ $G$ in \cite{der11}-\cite{vov12}, and $B>10^{-20} - 10^{-19}$ $G$ in \cite{fin15} and \cite{tak12}. At the same time, \cite{arl14} argued that null EGMF strength cannot be rejected with high statistical confidence. Most of cited values from \cite{der11}--\cite{arl14} were obtained assuming that the coherence length of the EGMF is about 1 $Mpc$ \cite{aka10}.

Some works (\cite{abr14},\cite{che15}) studied the angular patterns of the arriving photons in order to obtain the constraints. \cite{abr14} excluded the region $B = 3 \cdot 10^{-16} - 10{-14}$ $G$ and \cite{che15} found an indication for extended emission around some blazars, obtaining an estimate $B$=$10^{-17}$--$10^{-15}$ $G$. Finally, \cite{tas14} concluded that $B \approx 10^{-14}$ $G$; the uncertainty of this estimate is not clear. If the EGMF is stronger than $10^{-14}$ $G$ on the coherence scale 1 $Mpc$ then the cascade component in the spectra of point-like sources is almost entirely suppressed \cite{pro12}.

\section{Beyond the absorption-only model}

There are several inconclusive indications that the absorption-only model is incomplete \cite{che15}, \cite{hor12}--\cite{fur15} (for a more detailed discussion see \cite{dzh16}). According to \cite{hor12}, intrinsic spectra of some blazars reconstructed in the framework of the absorption-only model appear to have a pile-up in the optically thick regions (namely, these spectra are harder at high values of the $\gamma\gamma\rightarrow e^{+}e^{-}$ process optical depth $\tau>$2).

\begin{figure}[h]
\begin{minipage}{18pc}
\includegraphics[width=18pc]{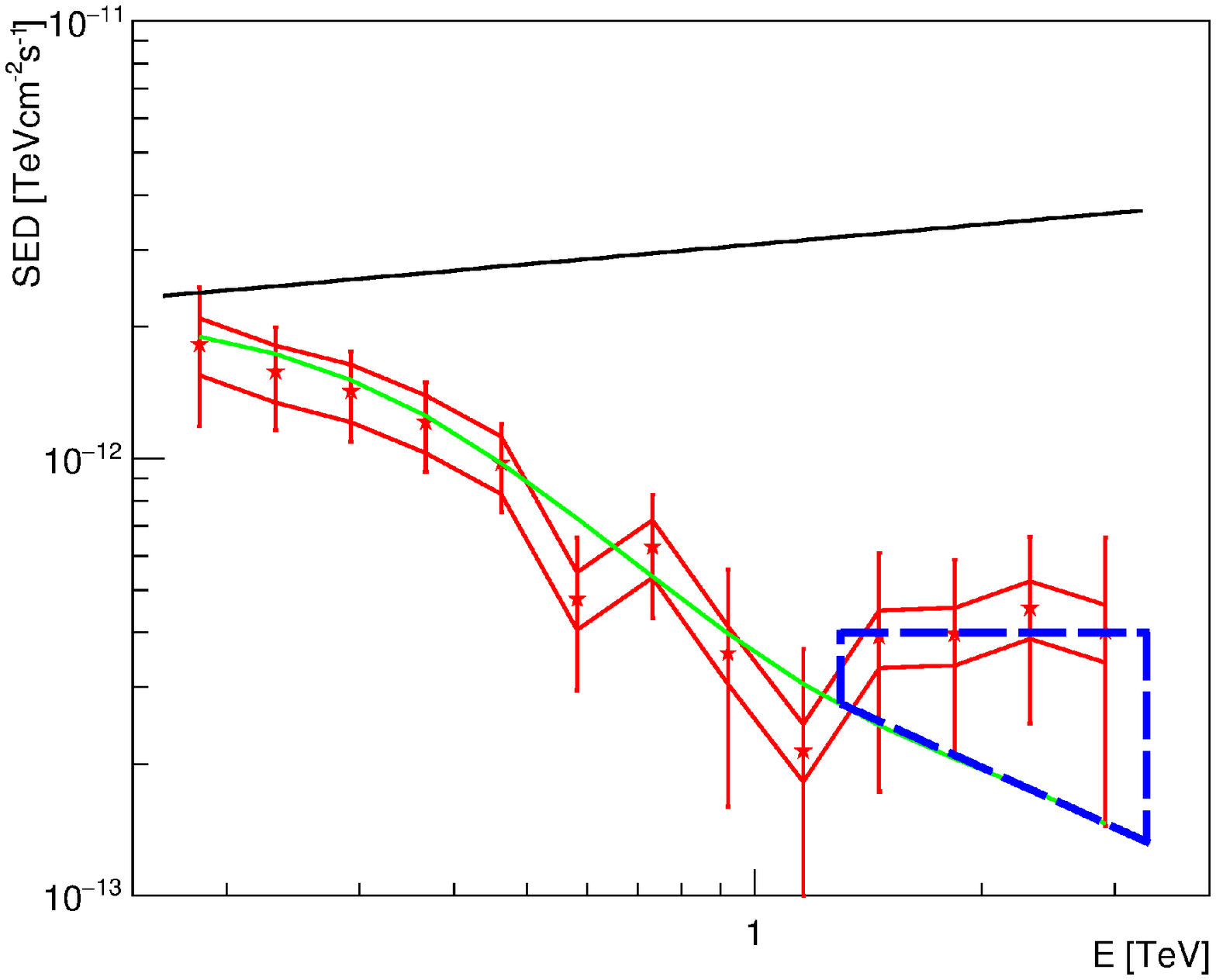}
\caption{\label{fig2} A fit to the SED of 1ES 1101-232 \cite{aha06} for the ``high-energy excess option'' (see text for more details). Red stars with bars show experimental data with statistical uncertainties, red broken lines denote systematic uncertainties. Black line is the intrinsic primary spectrum shape, green curve denotes the fit to the observable spectrum.}
\end{minipage}\hspace{2pc}%
\begin{minipage}{18pc}
\includegraphics[width=18pc]{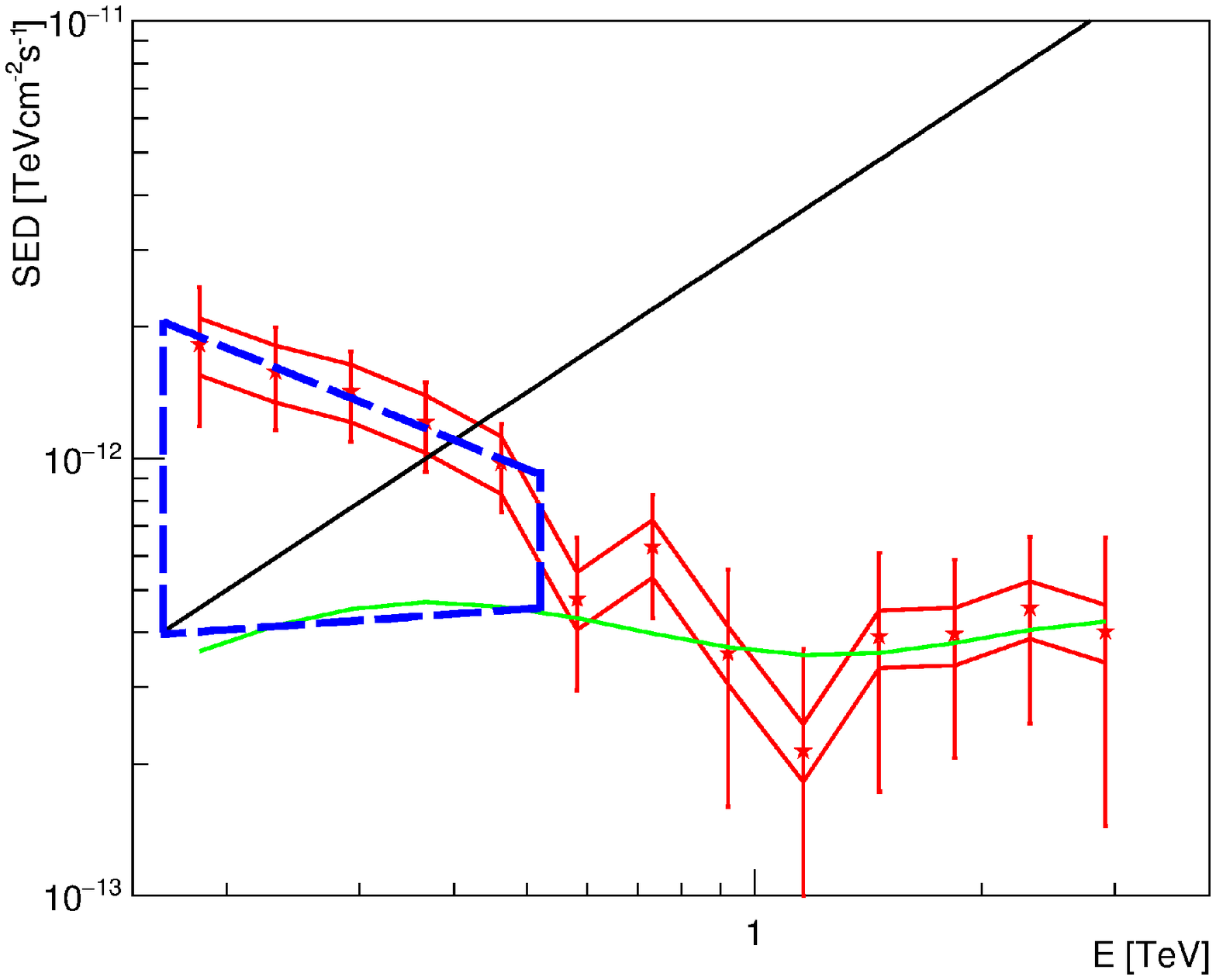}
\caption{\label{fig3} Another fit to the same SED of 1ES 1101-232 as in figure \ref{fig2}, but for the ``low-energy excess option''.}
\end{minipage} 
\end{figure}

This phenomenon may be explained assuming that {\it either} there is an excess at high energy, where $\tau>$2 (see figure \ref{fig2}) {\it or} an excess at some lower energy (see figure \ref{fig3}). In figure \ref{fig2} we present a fit to the spectral energy distribution (SED=$E^{2}dN/dE$, where $N$ is the number of observable $\gamma$-rays, and $E$ is energy) of blazar 1ES 1101-232 observed with the HESS detector \cite{aha06}. The fit was obtained assuming the absorption-only model and the EBL model of \cite{gil12}. A possible high-energy excess that is needed to explain the observable spectrum is denoted by dashed thick blue lines. Other option of a low-energy excess that also could explain the observed anomaly is presented in figure \ref{fig3}. The authors of \cite{hor12} implicitly assumed that the first option is realized, but, in fact the second one is not excluded {\it a priori}. In order to discriminate between them, one has to provide additional information.

Besides the spectral energy distribution of the source, contemporary Cherenkov imaging telescopes can also reconstruct the temporal and angular distributions of observable $\gamma$-rays. As was already mentioned, \cite{che15} found that some blazars have angular distribution of $GeV$ $\gamma$-rays broader than is expected for the case of a point-like source. Another study \cite{ner12} observed a low-energy cutoff at $E<$200 $GeV$ in the spectrum of Mkn 501 together with a possible broadening of an impulse shape towards lower energies. These works already indicate the presence of a low-energy excess, i.e. the {\it second} option considered above. As discussed in \cite{dzh16}, a plausible astrophysical interpretation of these phenomena is the presence of a cascade component in the observable spectrum.

\section{Electromagnetic cascade model}

\begin{figure}[h]
\includegraphics[width=17pc]{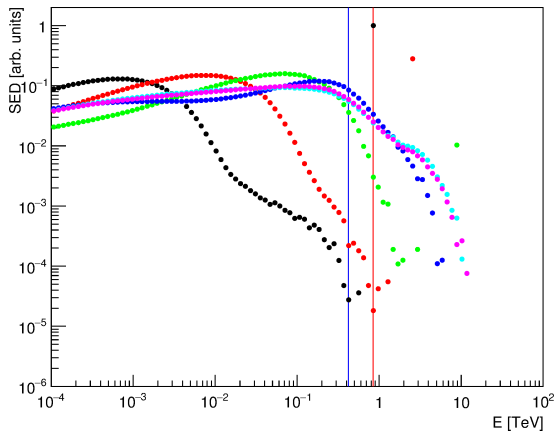}\hspace{2pc}%
\begin{minipage}[b]{19pc}\caption{\label{fig4} Observable gamma-ray spectra from primary monoenergetic gamma-rays of various energies. Black dots denote cascade spectrum with primary energy $E_{0}=$1 $TeV$, red - 3 $TeV$, green - 10 $TeV$, blue - 30 $TeV$, cyan - 100 $TeV$, and magenta - 1000 $TeV$. Singular dots denote primary spectra. Red vertical line is drawn at energy $E_{0}/(1+z)$, blue vertical line --- at $0.5\cdot E_{0}/(1+z)$. A similar figure was published in \cite{dzh16}.}
\end{minipage}
\end{figure}

Electromagnetic (EM) cascade model assumes that the primary particles of these cascades are $\gamma$-rays. In figure \ref{fig4} we present typical SEDs of cascade photons in the EM cascade model for the case of monoenergetic primary injection (primary energies range from 1 $TeV$ to 1 $PeV$). Calculations were performed with the publicly-available code ELMAG 2.02 \cite{kac12}. For the case of primary energy 1 $TeV$, two components of cascade photons may be identified --- comparatively low-energy $\gamma$-rays arising from inverse Compton (IC) interactions of cascade electrons on the CMB, and high-energy $\gamma$-rays produced by the same electrons on the EBL. This latter component produces a visible hardening of the observable spectrum above 10 $GeV$.

Two distinct regimes of cascade development are seen in this figure: the one-generation regime at primary energy $E_{0}<$10 $TeV$ and the universal regime at $E_{0}>$100 $TeV$. If the primary energy is comparatively low, the number of generations is small (1-2) and the spectrum becomes more dependent on both energy and type of the primary particle. In this case, the SED of the cascade photons has a distinct peak on energies roughly proportionate to the square of the primary particle energy. This scenario may be described by the one-generation approximation, hence its name the ``one-generation regime''. On the other hand, at high primary energy the observable spectrum is practically independent of the primary spectrum (for more details see \cite{ber16} and \cite{dzh16}). In \cite{dzh16} we showed that the electromagnetic cascade model, as a rule, allows to obtain good fits to observable spectra of considered blazars; cascade component mostly contributes at low energies, thus realizing the ``low-energy excess option'' shown in figure \ref{fig3}.

The third possible regime (not shown in the figure) is the ``extreme energy regime'', in which cascade electrons transfer most of their energy to the background photons in every act of the IC process (this regime may take place at ultrahigh energies of primary photons, $E_{0}>$1 $EeV$). We believe that further study of this regime could pave the way to a sensible and all-encompassing extragalactic gamma-ray propagation model.

\section{Hadronic cascade models}

In this section, we focus on the so-called hadronic cascade models, in which primary protons or nuclei create secondary electrons and photons by means of photohadronic processes on CMB and EBL and these particles, in turn, initiate EM cascades. It is particularly interesting to study hadronic cascade models for the case of ultra-high ($>10^{18}$ $eV$) and extremely high (EH) ($>10^{20}$ eV) primary proton energies since they could shed some light on the nature of the extragalactic Cosmic Rays (CR), e.g. help identify their sources. It is commonly known that the direct search for CR in the latter energy range is not feasible since both nuclei and protons rapidly lose their energy on the way: the former - on photodisintegration, the latter - on photopion losses. A more reasonable approach is to search the EH energy CR sources indirectly, e.g. search for secondary or cascade photons produced by the primary nuclei. The signatures of the hadronic cascade models are similar to the signatures of the EM cascade model in the universal regime, except much harder observable spectrum of the hadronic cascade model for $E>$200 $GeV$ (see \cite{dzh16}, section 2). Thus, the hadronic cascade model corresponds to the ``high-energy excess option'' shown in figure \ref{fig2}.

In \cite{dzh16} we showed that hadronic cascade models typically allow to obtain good fits of observable spectra for the sample of blazars considered here. However, turbulent magnetic fields that are believed to be present around CR sources may scatter primary nuclei, thus diminishing the effective observable flux. This is one of the main difficulties of hadronic cascade models that could be addressed in further studies.

\section{Conclusions}
In this paper we briefly reviewed the current status of $\gamma$-ray extragalactic propagation models. We showed that the strength of the Extragalactic Magnetic Field is poorly known, and the question whether the cascade component is present in the spectra of point-like sources is still open. Several observed effects suggest that this component may indeed contribute to the spectra of some blazars at sub-$TeV$ energies. There are indications that either a high energy excess or a low-energy excess in the spectra of these blazars should be present to explain the shape of the observed spectrum. Quite interestingly, we found that the simplest explanation of existing data in the $GeV$ and $TeV$ energy range is the electromagnetic cascade model that realizes the ``low-energy excess option''.

\subsection*{Acknowledgments}
This work was supported by the RFBR grant 16-32-00823.

\end{document}